\documentclass{nature}
\usepackage{amsmath}
\usepackage{amssymb}
\usepackage{hyperref}
\usepackage{lineno}
\usepackage{float}
\hypersetup{colorlinks,citecolor=blue,linkcolor=blue,urlcolor=blue}

\newcommand{\Fig}{Fig.}
\newcommand{\EFig}{Extended Data Fig.}

\newcommand{\src}{GLEAM-X\,J\,162759.5-523504.3}
\newcommand{\srcDM}{\ensuremath{57}}
\newcommand{\DMerror}{\ensuremath{1}}
\newcommand{\DMunit}{pc\,cm\ensuremath{^{-3}}}
\newcommand{\srcRM}{\ensuremath{-61}}
\newcommand{\RMerror}{\ensuremath{1}}
\newcommand{\RMunit}{rad\,m\ensuremath{^{-2}}}
\newcommand{\srcPlong}{\ensuremath{1091.1690}}
\newcommand{\srcPerr}{\ensuremath{0.0005}}
\newcommand{\srcP}{\ensuremath{1091}}
\newcommand{\srcalphaerr}{\ensuremath{0.04}}
\newcommand{\srcalpha}{\ensuremath{-1.16}}

\newcommand{\ndetections}{71}
\newcommand{\srcPdot}{\ensuremath{6\times10^{-10}}}
\newcommand{\Pdotunit}{\,s\,s\ensuremath{^{-1}}}
\newcommand{\ergpers}{erg\,s\ensuremath{^{-1}}}

\DeclareUnicodeCharacter{2212}{-}

\title{A radio transient with unusually slow periodic emission}

\author{N. Hurley-Walker$^{*1}$,
X.~Zhang$^{2,3}$, % contribution: polarisation calibration and analysis
A.~Bahramian$^{1}$, % contribution: X-ray and optical analysis
S.~J.~McSweeney$^{1}$, % contribution: dedispersion and barycentric corrections
T.~N.~O'Doherty$^{1}$, % contribution: archival search, initial discovery and second detection
P.~J.~Hancock$^{1}$, % contribution: methodology development, super-computing support
J.~S.~Morgan$^{1}$, % contribution: various aspects of the methodology including polarisation calibration
G.~E.~Anderson$^{1}$, % contribution: astrophysics analysis and  discussions
G.~H.~Heald$^{2}$, % contribution: polarization calibration
T.~J.~Galvin$^{1}$} % contribution: early refinement of period, polarization testing
\usepackage{graphicx}
\makeatletter
\let\saved@includegraphics\includegraphics
\AtBeginDocument{\let\includegraphics\saved@includegraphics}
\renewenvironment*{figure}{\@float{figure}}{\end@float}
\makeatother
\begin{document}

\maketitle

\begin{affiliations}
 \item International Centre for Radio Astronomy Research, Curtin University, Kent St, Bentley WA 6102, Australia
 \item CSIRO, Space and Astronomy, PO Box 1130, Bentley WA 6102, Australia
 \item Shanghai Astronomical Observatory, Chinese Academy of Sciences, 80 Nandan Road, Shanghai 200030, China
\end{affiliations}

\begin{abstract}

The high-frequency radio sky is bursting with synchrotron transients from massive stellar explosions and accretion events, but the low-frequency radio sky has so far been quiet beyond the Galactic pulsar population and the long-term scintillation of AGN. The low-frequency band, however, is sensitive to exotic coherent and polarised radio emission processes such as electron cyclotron maser emission from flaring M-dwarfs\cite{2017ApJ...836L..30L}, stellar magnetospheric plasma interactions with exoplanets\cite{2020NatAs...4..577V}, and a population of steep-spectrum pulsars\cite{2021ApJ...911L..26S}, making Galactic plane searches a prospect for blind transient discovery. Here we report an analysis of archival low-frequency radio data that reveals a periodic, low-frequency radio transient. We find that the source pulses every 18.18\,minutes, an unusual periodicity not previously observed. The emission is highly linearly polarised, bright, persists for 30--60 s on each occurrence, and is visible across a broad frequency range. At times the pulses comprise short ($<0.5$-s) duration bursts; at others, a smoother profile is observed. These profiles evolve on timescales of hours. By measuring the dispersion of the radio pulses with respect to frequency, we have localised the source to within our own Galaxy, and suggest that it could be an ultra-long period magnetar.

\end{abstract}

\begin{figure}[H]
\includegraphics[width=0.4\textwidth]{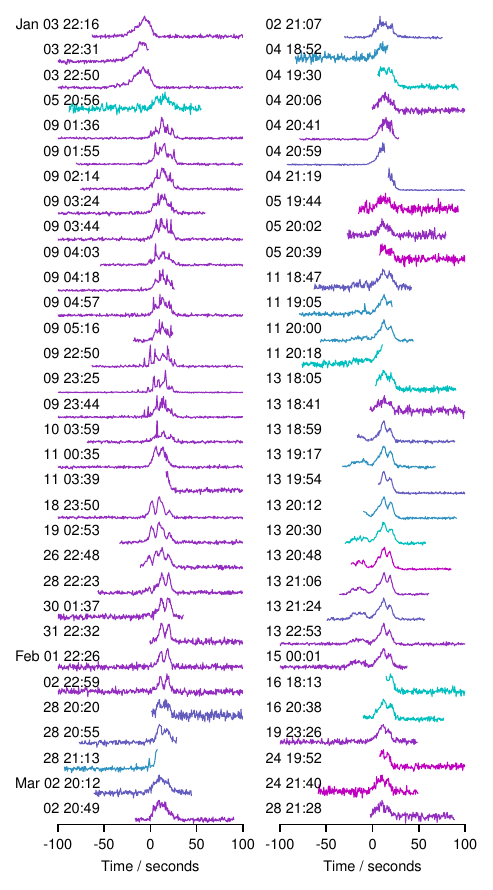}
\caption{64 of the \ndetections{} detected pulses of \src{} aligned by its measured period $P$ and period derivative $\dot{P}$. Omitted pulses were truncated or too low in signal-to-noise to be displayed here; flux densities are normalised to the peak of each pulse for readability; barycentric and dispersive corrections have been applied. The observation start times in UTC are listed on the left of each detection. The color range spans 88\,MHz (cyan) to 215\,MHz (magenta) and the detections span 84~days. The pulses observed on Jan~03 are not misaligned and fit within the widest pulse windows found in Mar~13 (see Methods).\label{fig:light_curves}}
\end{figure}

\begin{figure}[H]
\includegraphics[width=\textwidth]{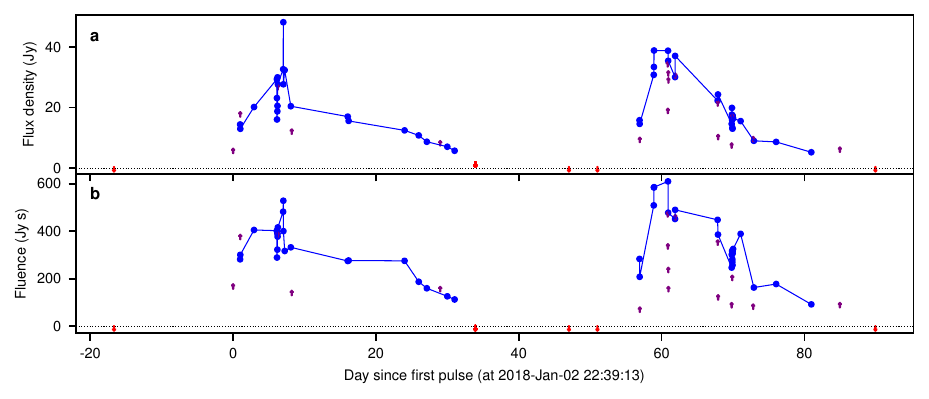}
\caption{Maximum brightness of each pulse (a) and total pulse fluence (b) as a function of time across the two observed intervals of activity. Not all pulses are fully captured by every observation; in these cases lower limits are plotted with purple upward-pointing arrows. Observations in which the source was within the field-of-view and predicted to be detectable, but not found, are shown with red downward-pointing arrows equal to the root-mean-square noise in a 30-s image corresponding to the time we would have expected a pulse. All measurements have been scaled to a common frequency of 154\,MHz via the spectral index $\alpha=\srcalpha$. Error bars are omitted for clarity, and are dominated by the $\sim5$\,\% uncertainty in the primary beam model of the telescope.\label{fig:fluence}}
\end{figure}

We searched 24~hours of Galactic Plane observations taken by the Murchison Widefield Array (MWA; see Methods) using a rapid shallow search method to probe a novel transient timescale parameter space (Hancock et al. in prep). From this search, we discovered the transient source \src{}, initially with two significant detections. An exhaustive further search yielded \ndetections{} pulses spanning January to March 2018 (\Fig~\ref{fig:light_curves}), with maximum flux densities ranging from 5--40\,Jy (\Fig~\ref{fig:fluence}). The pulse widths range between 30 and 60\,s, and evolve on hourly timescales, sometimes comprising many ``spiky'' bursts unresolved at our time resolution of 0.5\,s, other times displaying sub-pulses of widths 10--30\,s. Aligning the pulses, we established 
a period of \srcPlong{}$\pm$\srcPerr{}\,seconds (see Methods). It was necessary to perform a barycentric correction, indicating an extrasolar origin.

Using data spanning 72--231\,MHz, we established a dispersion measure of \srcDM$\pm$\DMerror\,\DMunit{} (see Methods; \EFig~\ref{fig:dm}), which when combined with Galactic electron density models\cite{2017ApJ...835...29Y}, produces a distance estimate of $1.3\pm0.5$\,kpc. We also measured a radio spectral index of $\alpha=\srcalpha\pm\srcalphaerr$, where the flux density $S_\nu \propto \nu^\alpha$, indicating that the emission is non-thermal. Brightness variations on timescales of less than 0.5\,s imply a compact object, and the brightness temperature of a thermal object 0.5\,light~seconds in size producing 20\,Jy of flux density at 1.3\,kpc is $\approx10^{16}$\,K, implying a coherent emission mechanism. 

\begin{figure}[H]
\includegraphics[width=\textwidth]{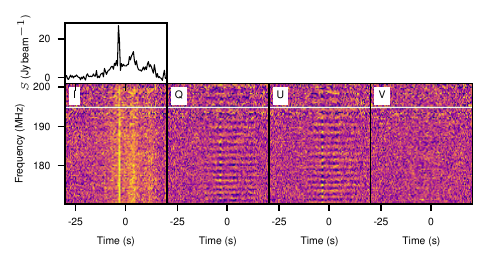}
\caption{Dynamic spectrum of the observation recorded at 2018-01-10 03:59. From left to right the panels show the Stokes I, Q, U, and V flux density as a function of frequency and time, with a dispersion correction of \srcDM{}\,\DMunit{} applied. Linear Stokes Q and U show Faraday rotation of \srcRM$\pm$\RMerror\,\RMunit{} while circular V shows no obvious signal. The top left panel of the image shows the profile of the Stokes~I data averaged over the frequency axis; the unresolved burst of emission shows the limitation of our 0.5~s time resolution. The root-mean-square of the noise in each spectrum is $8.5$\,Jy\,beam$^{-1}$, and $0.9$\,Jy\,beam$^{-1}$ in the summed profile.\label{fig:dynspec}}
\end{figure}

The pulses exhibit a constant, high fractional linear polarisation ($88\pm1$\,\%) and there is no change in polarisation angle as either a function of pulse phase or observation time (Fig.~\ref{fig:dynspec}). The Faraday rotation measure (RM) remains constant over the observations at \srcRM$\pm$\RMerror\,\RMunit, which is consistent with the Galactic RM towards this region\cite{2020A&A...633A.150H} ($-72.3\pm154.9$\,\RMunit{}).

The fluence of the pulses is variable but broadly follows a distribution with two ``on'' intervals $\sim30$~days wide (Jan~03 -- Feb~02 and Feb~28 -- Mar~28: see \Fig~\ref{fig:fluence}), with fast rise times, slow decay times, and a 26-day null interval between them. The detections are all serendipitous in archival data, leading to heterogeneous coverage. During the ``on'' intervals, at every time where we predict we would detect a pulse, we do so (i.e. there is no obvious nulling). The flux density in 30-s images at the source location is $<10$\,mJy during the sampled ``off'' intervals (upper limits in \Fig~\ref{fig:fluence}). While the fluence variation does somewhat resemble gravitational lensing\cite{2017ApJ...845...89V} or extreme scattering events\cite{2016Sci...351..354B}, the magnification required would be of order 1,000$\times$, and difficult to create with any physically plausible gravitational or plasma lens.

The high linear polarisation indicates the presence of strongly ordered magnetic fields; this and the luminosity of the pulses are not explicable by known phenomena such as radio emission from flare stars\cite{2010ARA&A..48..241B}, exoplanets\cite{2001Ap&SS.277..293Z}, and white dwarf/m-dwarf binaries\cite{2016Natur.537..374M}, all of which would be orders of magnitude fainter at this distance, and typically circularly polarised.
The extreme regularity of the emission (fractional uncertainty $\frac{\sigma_P}{P}<5\times10^{-7}$) implies either a rotational or orbital origin. Due to the $0.5$\,light-second upper limit on the object's size, a rotational origin may be more likely.

\begin{figure}[H]
\includegraphics[width=\textwidth]{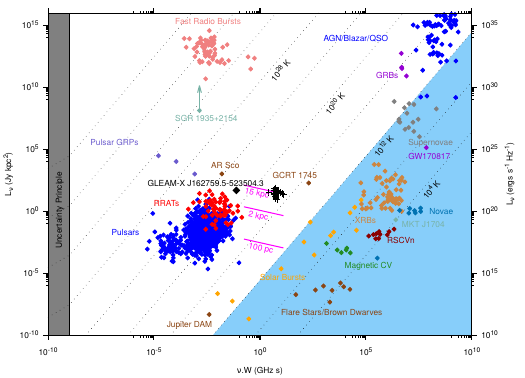}
\caption{Transient parameter space populated with common known radio transients. Radio pseudo-luminosity is shown on the vertical axis and the product of observing frequency $\nu$ and transient/variability timescale $W$ is plotted on the horizontal axis. The shaded blue region shows objects with brightness temperatures that do not imply coherent emission mechanisms. Black ``+'' marks show the $W=30$--60-s pulses of \src{}, while a black filled diamond indicates a representative unresolved pulse of $W=0.5$\,s and $S=30$\,Jy. Diagonal magenta lines show the sensitivity of the MWA to long-pulse transients at Galactic distances, for $W=2$--60\,s across 72--231\,MHz, with dispersion measure smearing effects included. This figure is adapted from Pietka et al. 2015\cite{2015MNRAS.446.3687P}.
\label{fig:LnuW}}
\end{figure}

\Fig~\ref{fig:LnuW} shows \src{} in context with other sources of transient emission. The source that bears the most similarity to \src{} is GCRT\,1745, a transient radio source detected toward the Galactic Centre with observations at 330\,MHz, which exhibited five 10-minute duration bursts with periodicity 77\,min\cite{2005Natur.434...50H}. These bursts exhibited slow profile evolution and low circular polarisation (no linear polarisation measurements were available). Due to the crowded field in this direction, and lack of distance constraints, Hyman et al. were unable to determine a progenitor but speculated that it may be a long-period magnetar, an interpretation that can be applied to \src{}.

Magnetic neutron stars rotating at period $P$ lose energy via magnetic dipole radiation, causing them to spin down (i.e. they have a positive period derivative $\dot{P}$); this spin down luminosity ($\dot{E}$) can be expressed as $\frac{4 \pi I \dot{P}}{P^3}$, where $I$ is the neutron star moment of inertia, typically assumed to be $10^{45}$\,g\,cm$^{-2}$.
Via a grid search for $P$ and $\dot{P}$ (see Methods), we find that $\dot{P}>0$ is preferred, the best fit value is $\dot{P}=\srcPdot{}$\Pdotunit{}, the analysis favours $\dot{P}<1.2\times10^{-9}$\Pdotunit{} (see \EFig~\ref{fig:p_pdot} for context with known neutron stars). Using this value as an upper limit, we find a maximum spin-down luminosity of $\dot{E}<1.2\times10^{28}$\,\ergpers. The flux density of radio pulsars can be converted to a luminosity (see Methods) which for the brightest pulses from \src{} is $4\times10^{31}$\,\ergpers. 
Pulsar radio luminosities are typically a small fraction of their spin-down luminosities\cite{2014ApJ...784...59S}, while this source exhibits the reverse, indicating that the emission is not generated purely by spin-down. Additionally, the smooth variations in pulse profile and the transient window of radio emission are more consistent with the interpretation of \src{} as a radio magnetar than a pulsar\cite{2012MNRAS.422.2489L}.

Magnetars are commonly detected and characterised via X-ray observations; four of the five known magnetars that have produced detectable pulsed radio emission have done so only after X-ray outbursts. However, not all X-ray emitting magnetars produce detectable radio emission. Previous studies have shown that magnetars only produce radio emission if their quiescent X-ray luminosity in the 0.5--10\,keV band is lower than their spin-down luminosity\cite{2012ApJ...748L..12R}. We would therefore predict that the X-ray luminosity $L_\mathrm{X}$ of \src{} is $<6\times10^{27}$\,erg\,s$^{-1}$. We obtained X-ray observations with the \textit{Swift}/XRT and determined that $L_\mathrm{X}<10^{32}$\,erg\,s$^{-1}$ (see Methods), which while not a strong limit compared to our expectation under this interpretation, is a lower quiescent X-ray luminosity than all but two of the faintest known magnetars\cite{2014ApJS..212....6O}, SGR\,$0418+5729$ and Swift~J\,$1822.3-1606$. Alternatively, a white dwarf would have a moment of inertia and therefore spin-down luminosity $10^5\times$ larger, allowing the possibility of spin-powered radio pulsations; deeper UV observations than currently available would test this hypothesis.

Regardless of interpretation, the existence of an unexpected slowly pulsating, yet intermittent radio transient opens up an entirely new field of exploration of radio surveys, particularly at low frequencies.
While many sensitive low-frequency ($\lesssim340$\,MHz) continuum surveys have searched for transients on cadences of minutes within extragalactic fields for up to an hour at a time, no such systematic survey for unknown minute-period transients has been conducted within the Galactic plane on similar timescales\cite{2016MNRAS.459.3161C,2016ApJ...832...60P,2016MNRAS.456.2321S,2019MNRAS.482.2484B,2019MNRAS.490.4898H}. Since known pulsars and magnetars have periods of $\lesssim 10\,$s, surveys are typically designed with relatively short dwell times of $\approx100$--5000\,s\cite{Keith2010}
and employ high pass filters to mitigate red telescope noise\cite{2017MNRAS.468.1994C}, rendering them insensitive to long-period sources.
Current MWA survey strategies\cite{2017MNRAS.464.1146H,2019MNRAS.482.2484B} may therefore be sensitive to a large population of these sources and further detections may explain some of the other puzzling unidentified low-frequency Galactic transients\cite{2005Natur.434...50H}.

We can estimate the population of similar long-period transients detectable with future searches starting from the following assumptions: like other first detections of a new class of object, \src{} is the most luminous example of its kind; the population's luminosity and spatial distribution are similar to that of pulsars; other sources would have similar long-term duty cycles of $\approx2$\,\% (see Methods). Now that we are aware of this class of objects, it is worth the computational expense of performing model-subtracting time-step imaging (see Methods), and searching all voxels for significant residuals, a technique $10\times$ more sensitive than the fast search that detected \src{} (see diagonal magenta lines in \Fig~\ref{fig:LnuW}). With these population assumptions, and using this technique, we predict $\approx10$~further detections of similar objects in the same volume of data that yielded \src{} (24~hours). Larger duty cycles or a flatter luminosity distribution would increase this number. The MWA archive contains thousands of hours of observing time sensitive to $|b|<10^\circ$, which may potentially yield hundreds of further similar objects; this will also be true of other radio archives with data covering the Galactic Plane. Further detections and rapid high-time-resolution follow-up of candidates will more conclusively determine the nature of these sources, thus providing further insight into the evolutionary extremes surrounding the life and death of massive stars. 

%\include{supplementary}
% Used this up until the time for final submission, then did the final compilation and downloaded the bbl file, and replaced this line.
%\bibliography{refs.bib}

\begin{methods}

\subsection{Observations}
The Murchison Widefield Array is a low-frequency radio telescope operating in the Murchison region of Western Australia\cite{2013PASA...30....7T, 2018PASA...35...33W}. At the time of the observation in which \src{} was discovered, it was engaged in observing the GaLactic and Extragalactic All-sky MWA -- eXtended (GLEAM-X) survey, a follow-up to the GLEAM survey\cite{2015PASA...32...25W}. The data were taken in a drift scan mode, iterating through a 72--231\,MHz bandwidth by dwelling for 2\,minutes in each of five 30.72-MHz bands, yielding a resolution of $5'$--$45''$ and snapshot noise levels of $150$--$25$\,mJy\,beam$^{-1}$. Over this bandwidth the field-of-view of the instrument is $60^\circ$--$25^\circ$ across, yielding multiple measurements of sources drifting through the primary beam. The data are sampled at 0.5-s time resolution and 10-kHz frequency resolution. We downloaded measurement sets from the MWA All-Sky Virtual Observatory (https://asvo.mwatelescope.org/) and at this stage averaged the data to 40\,kHz to reduce data size and decrease processing times. 

The initial detection of the transient was made by performing a differencing of the visibilities between observations taken at an identical Local Sidereal Time several months apart (Hancock et al. in prep). Two detections were made, and based on the non-detections in adjacent observations, an initial period was determined. Including a barycentric correction, and searching the archive thoroughly between July~2017 and July~2018, the interval of activity was established to be January~2018 to March~2018, with a total of \ndetections{}~detections (see Supplementary Methods for further details on searching the MWA archive).

\subsection{Calibration and imaging}
For all data, calibration was performed using the \textsc{mitchcal} algorithm\cite{2016MNRAS.458.1057O} to derive antenna gains by comparing the raw data to model visibilities formed from a sky model based on catalogues derived from GLEAM\cite{2017MNRAS.464.1146H, 2019PASA...36...47H}. These were applied to the observation, and deep imaging was subsequently performed using \textsc{WSClean}\cite{2014MNRAS.444..606O} version~2.9.0, creating a visibility model that includes the spectrally variant primary beam. A $10'\times10'$ region around \src{} was masked during this process, so that the source is not included in the model. 

After the deep model was created and subtracted from the visibilities, the data were re-imaged such that the transient was the only source in the field. The data were imaged at 320\,kHz frequency resolution and 0.5\,s time resolution in full Stokes (I,Q,U,V), which were formed from the instrumental data by application of the MWA beam model\cite{2015RaSc...50...52S}. We also used observations of a polarisation calibrator (PKS~J0636$-2036$) to correct radio emission leaking from Stokes~U into Stokes~V (X-Y phase calibration\cite{2017PASA...34...40L}).

\subsection{Dispersion Measure}
Since strong pulse profile evolution is observed, we used the five observations that spanned 72--231 MHz in the shortest interval of time (74~minutes; \EFig~\ref{fig:dm}) to calculate the dispersion measure.
We aligned these observations using a period of \srcPlong\,s, then ran dedispersion trials in steps of 0.5 over range 50 to 60\,\DMunit, finding that $\srcDM\pm\DMerror{}$\,\DMunit{} produced the best fit. Light curves (\Fig~\ref{fig:light_curves})
were then produced for each observation by applying dedispersion and averaging the dynamic spectra along the frequency axis.

\subsection{Polarisation}
Polarisation analysis was performed following the methods of the POlarised GLEAM Survey (POGS\cite{2018PASA...35...43R,2020PASA...37...29R}). From all observations with detections, only observations within 170--230 MHz were selected for the polarisation analysis to avoid depolarisation caused by the (40\,kHz) channel width. We performed RM synthesis\cite{brentjens2005faraday} on the time-averaged Q/U spectrum of each observation with the RM Tools software (\href{https://github.com/CIRADA-Tools/RM-Tools}{https://github.com/CIRADA-Tools/RM-Tools} to obtain the corresponding RM, fractional polarisation, and polarisation angle (see \Fig~\ref{fig:iquvprm} for these results in context with the surrounding region of sky). To investigate any variation of the polarisation angle within the pulse phase, we also performed RM synthesis on the Q/U spectrum of each timestep for two high S/N observations. We found that the polarisation angle was constant with respect to time within and between observations (e.g. the middle two panels of \Fig~\ref{fig:dynspec}).

\subsection{Period and Period Derivative}

We used phase dispersion minimization based on Lafler-Kinman statistic\cite{1965ApJS...11..216L,1978ApJ...224..953S,1996ApJ...460L.107S,2002A&A...386..763C} to quantify the periodicity of the pulses, using the package \textsc{p4j} (\url{https://github.com/phuijse/P4J}). This choice is motivated by the variable shape and amplitude of the pulses and presence of large irregular gaps in data (which occasionally truncate part of an observed pulse). Using this method we found a clear peak in the periodogram with a period of $1091.170$\,s.

The period obtained from the periodogram analysis above assumes that the period is constant (i.e. $\dot{P} = 0$).
In order to place constraints on $\dot{P}$, we performed a grid search of $P$ and $\dot{P}$ values, centred on $1091.170$\,s and $0$\,\Pdotunit{} respectively, to find which pairs of values are consistent with the observed arrival times of the pulses.
We searched over the ranges $1091.150$\,s$\le P \le 1091.185$\,s and $-4\times10^{-9} \le \dot{P} \le 4\times10^{-9}$, aligning the pulses in pulse phase for each value pair, and taking the peak flux density of the averaged pulses (the ``mean profile'' in pulsar parlance) as our metric for goodness of alignment.
Since different pulses were observed in different frequency bands, each brought to a common frequency of 154\,MHz using $\alpha=\srcalpha{}$ (see below), before averaging.

As shown in \EFig~\ref{fig:ppdot}, the resulting $P$ and $\dot{P}$ were somewhat degenerate, as expected due to the three-month time window of measurements and the evolving pulse profile. The maximum peak flux density 15.5\,Jy corresponds to the values $P = \srcPlong{}\pm\srcPerr{}$\,s and $\dot{P} = \srcPdot{}$\Pdotunit{}. The most significant contour in \EFig~\ref{fig:ppdot} is shown at 15\,Jy to show that there are multiple values of $P$ and $\dot{P}$ which produce a similar flux density. Therefore, this analysis favours the ranges $0 < \dot{P} < 1.2\times10^{-9}$, i.e. spin-down is more likely than spin-up. Note that the error on the source period is representative for any choice of $\dot{P}$ within this range, reflecting the width of the most significant contour in the period dimension.

\subsection{Spectral index}
\label{sec:spectral_index}

Individual 30.72-MHz observations of \src{} lack the frequency coverage to determine an accurate spectral index, but since the flux density clearly varies with time, averaging over many observations is unlikely to yield a usable result. To obtain a good estimate, we used the same observations with which we measured the dispersion measure: 1205008192 (72--103\,MHz), 1205007112 (103--134\,MHz), 1205011432 (139--170\,MHz), 1205010352 (170--200\,MHz), and 1205009272 (200-231\,MHz); these are the observations taken on 2018-03-13 between 20:12 and 21:24 (\Fig~\ref{fig:light_curves}), where the pulse profile is reasonably consistent with time. After dedispersion and alignment on the source period, we obtained an average profile by averaging the data along the combined frequency axis (vertical axis in the right-hand panel of \EFig~\ref{fig:dm}). From this we selected the timesteps where the source was clearly ``on'' (timesteps 0--100 in the right-hand panel of \EFig~\ref{fig:dm}), and for each frequency, determined the weighted average flux density, using the average profile as a weighting function.

Errors were calculated as the root-mean-squared noise of each slice for unflagged timesteps during which the source was ``off'', divided by the square root of the number of samples in the average flux density measurement, added in quadrature with a 5\,\% flux density calibration error. \EFig~\ref{fig:spectrum} shows the weighted averages and their errors plotted as a function of frequency. The drop-off at low frequencies is likely due to averaging over the strong ionospheric scintillation (time-dependent striping visible in the top of \EFig~\ref{fig:dm}), while the steepening at the higher frequencies may be intrinsic, but the signal-to-noise is low so it is difficult to be sure. Therefore, we used the measurements taken between 95 and 195\,MHz and the \texttt{scipy} implementation of a the Levenberg-Marquardt least-squares fit to determine a radio spectral index of $\alpha=\srcalpha\pm\srcalphaerr$, where the flux density $S_\nu \propto \nu^\alpha$. The reduced-$\chi^2$ of the fit is 1.86.

\subsection{Error on the distance estimate}

Models of the Galactic electron density can be used to convert the dispersion measure (DM) into a distance. YMW~2016, the most recent model\cite{2017ApJ...835...29Y}, derives a relative distance error $\frac{D_m - D_i}{D_i}$ using 189~pulsars for which distances are known: $D_m$ is the model distance based on the observed DM and $D_i$ is the independently determined distance (e.g. by parallax). For their sample, the root-mean-square of the relative distance error is 0.398. From version 1.65 of the Australia Telescope National Facility Pulsar Catalogue\cite{2005AJ....129.1993M}, we extracted data for the five pulsars with independently-measured distances within $10^\circ$ of \src{}. For these pulsars, we found that the root-mean-square of the relative distance error is 0.393. Thus we conclude that there is a $\sim40$\,\% error on the estimated distance of 1.3\,kpc.

\subsection{Position measurement}
At radio frequencies, the apparent source position measured by an interferometer may be shifted by an angular offset $\Delta \theta$ by the ionosphere, proportional to the transverse gradient $\nabla_{\perp}$ of the total electron content (TEC) toward the source and the square of the wavelength at which the observations were taken.
% Removed \cite{TMS} because we have already hit the reference limit
To determine an accurate position for \src{}, we used observation 1205009272, taken at the highest frequency band, 200 -- 231\,MHz, on a quiet night during which ionospheric distortions were minimised, and imaging just the 30-s subset of the observation during a high S/N pulse. We used the software \textsc{fits\_warp}\cite{2018A&C....25...94H} to calculate the local position shift based on the apparent shifts of nearby compact bright calibrator sources. The error was calculated as the mean residual offset of these nearby sources after the shift had been modelled and removed, i.e. $2''$. The derived position is R.A. $16^\mathrm{h}27^\mathrm{m}59.5^\mathrm{s}$, Dec. $-52^\circ35'04.3''$.

%\begin{equation}
%\Delta \theta = - \frac{1}{8\pi^2}\frac{e^2}{\eta_0 m_e } %\frac{1}{\nu^2} \nabla_\perp \mathrm{TEC}
%\label{eq:d_theta}
%\end{equation}

\subsection{\textit{Swift}/XRT Observations}
We observed \src{} with the Neil Gehrels {\it Swift} observatory\cite{2004ApJ...611.1005G} for 2\,ks with the X-ray telescope (XRT)\cite{2005SSRv..120..165B} in photon-counting mode.
%and the UV/optical telescope (UVOT)\cite{2005SSRv..120...95R} with uniform weighting in all 6 filters.
The observations were taken from 2021-02-25 16:34:00 to 2021-02-25 21:27:00 (UTC) and the observation id was 00014085001.
We reduced and analyzed the data using HEASOFT~v6.28\cite{2014ascl.soft08004N}, and the \textsc{xrtpipeline}~v0.13.5. 
There is no detection of a point source in any bands, nor is there any indication of extended structure.
We estimate a 3$-\sigma$ upper limit of 5.6$\times10^{-3}$\,ct\,s$^{-1}$ for the count rate in the 0.3--10\,keV band for \src{}. Assuming a thermal spectrum (blackbody, $kT=0.1$\,keV), the upper limit in count rate corresponds to an absorbed flux of 1.5$\times10^{-13}$ \ergpers\,cm$^{-2}$ in the 0.3--10\,keV band.
Assuming a non-thermal spectrum ($E^{-\alpha}$; $\alpha=2$, typical for magnetars\cite{2014ApJS..212....6O}), the upper limit in count rate corresponds to an absorbed flux of 1.9$\times10^{-13}$\,\ergpers\,cm$^{-2}$ in the 0.3--10\,keV band. \EFig~\ref{fig:xray_magnetars} shows the corresponding luminosity limit calculated as a function of $kT$.

\subsection{Radio luminosity calculation}

Working under the assumption that \src{} is a pulsar or magnetar in order to determine its energetics, we can transform the observed flux density into a radio luminosity.
To precisely determine the radio luminosity of a rotating magnetic neutron star its geometry with respect to the observer needs to be known; this is often derived by examining the change in pulse phase with respect to time.
In the case of \src{}, the pulse phase is flat, similar to the case of the radio magnetar Swift~J\,$1818.0-1607$\cite{2020MNRAS.498.6044C}. In this case, we cannot derive the geometry of the emission cone
%\cite{2021MNRAS.502.1549T}
and instead interpret the flat phase as our line-of-sight just grazing the edge of the emission cone (i.e. the impact angle between the line-of-sight and the magnetic axis is similar to the emission cone opening angle). This is qualitatively consistent with the pulse duty cycle of 30--60\,s of activity in \srcP{}\,s, $\approx3$--6\,\%. For pulsars with this duty cycle, for typical opening angles of $6^\circ$, the radio luminosity at 1.4\,GHz is\cite{2012hpa..book.....L} $7.4\times10^{30}\frac{D}{\mathrm{kpc}}^2\frac{S_\mathrm{1.4GHz}}{\mathrm{Jy}}$\,\ergpers. \src{} produces pulses of peak flux densities of up to $S_\mathrm{154MHz}=45$\,Jy. Scaling this to 1.4\,GHz by $\alpha=\srcalpha{}$ we would expect $S_\mathrm{1.4GHz}=3.5$\,Jy, and therefore $L_\mathrm{1.4GHz}=4\times10^{31}$\ergpers.

\subsection{Long-term duty cycle}
In 8~years of operation, the MWA has accumulated $\sim160$\,hours of observing time that have pointing directions within 15$^\circ$ of \src{}, and might in principle be sensitive to it. However, since the data span many different projects, array configurations, frequencies, and observing modes, and processing data takes about $100\times$ longer than observing, searching this data thoroughly is a daunting task. Additionally, only when the data are taken in a contiguous $\sim$20-minute block may emission from this source be ruled out. We first examined the data before the 2018-Jan -- Mar activity window using the period of the source, and found no emission from \src{} in 2017-Dec or 2018-Apr. We searched five blocks of 20-minute contiguous observations in 2017-May, 2017-Oct, 2017-Nov, 2018-Apr, 2018-May, finding no detections. Before 2017-May and after 2018-Jun, the telescope was reconfigured into compact ``Hex'' mode, a redundant-baseline configuration with very poor imaging quality, yielding poor constraints from data taken in that mode. We searched for suitable observations in the archive as early as possible (2014-Mar and 2014-Jun), and also took new observations (2021-Feb), finding no detections. There remains in the archive $\sim15$\,hours of data in suitable contiguous 20-minute blocks that may be searched in a future paper.

At the time of writing, the MWA has been in operation for 8~years, and this source was only found to be active for 2~months in that time, yielding an estimate of the duty cycle of $\frac{60\,\text{days active}}{3000\,\text{days searched}} = 2$\,\%. Alternatively, we could assume that the source was active any time that we have not thoroughly searched the data, and with the co-operation of a conspiratorial Universe, the duty cycle can trend toward 100\,\%. Or, this might be the only time in the (unknown) lifetime of the source that it has produced emission, resulting in a duty cycle trending toward 0\,\%. It is also worth noting that the source may not be entirely inactive during our non-detections, but its pulsations may be below our detection threshold; follow-up with a more sensitive radio telescope would be illuminating. For this work, we use the estimate of a $2$\,\% long-term duty cycle for pulses above the detection threshold of the MWA.

\subsection{Pulse profile evolution and alignment}

At first examination, the three pulses recorded on Jan~03 appear to be misaligned with the other pulses. However, the pulses recorded on Mar~13 show an early ``secondary'' pulse preceeding the main pulse by $\sim15$\,s. Aligned by our determined $P$ and $\dot{P}$, the pulses from Jan~03 fit within the wide pulse profiles found in Mar~13 (\EFig~\ref{fig:pulse_alignment}). We therefore suggest that there is no misalignment, and instead the apparent effect is caused by pulse profile evolution.

\end{methods}

\begin{addendum}
 %\item We thank the anonymous referee for their comments, which improved the quality of this paper.
 \item N.H.-W. is the recipient of an Australian Research Council Future Fellowship (project number FT190100231) and G.E.A. is the recipient of an Australian Research Council Discovery Early Career Researcher Award (project number DE180100346) funded by the Australian Government. This scientific work makes use of the Murchison Radio-astronomy Observatory, operated by CSIRO. We acknowledge the Wajarri Yamatji people as the traditional owners of the Observatory site. Support for the operation of the MWA is provided by the Australian Government (NCRIS), under a contract to Curtin University administered by Astronomy Australia Limited. We acknowledge the Pawsey Supercomputing Centre which is supported by the Western Australian and Australian Governments. The National Radio Astronomy Observatory is a facility of the National Science Foundation operated under cooperative agreement by Associated Universities, Inc. This project was supported by resources and expertise provided by CSIRO IMT Scientific Computing. This work used resources of China SKA Regional Centre prototype funded by the National Key R\&D Programme of China (2018YFA0404603) and Chinese Academy of Sciences (114231KYSB20170003).
 \item[Author Contributions] N.H.-W. calibrated and processed the data for the observations described herein, determined the position and flux density of the source, and prepared the manuscript with contributions from all co-authors. X.Z. processed all polarization data, including performing polarization calibration and analysis. A.B. and S.J.M. performed the analysis to derive the period and period derivative. A.B. performed the X-ray observations and analysis. S.J.M. calculated and applied the dispersion measure and barycentric corrections. T.N.O. developed the original detection methodology, performed the original archive search, and made the initial discovery. P.J.H. helped develop the detection methodology and provided super-computing support. G.E.A. contributed to astrophysical calculations and interpretation of the data. T.J.G., G.H.H., J.S.M., and X.Z. determined polarization calibration methods. T.J.G. performed early refinement of the period estimate.
 \item[Competing Interests] The authors declare that they have no
competing financial interests.
 \item[Data Availability]
 Data that supports this paper are available at the following public repository: \newline\href{https://github.com/nhurleywalker/GLEAM-X\_Periodic\_Transient}{https://github.com/nhurleywalker/GLEAM-X\_Periodic\_Transient}. Further data products can be supplied by the authors on reasonable request.
  \item[Code Availability] Code that supports this paper is available at the following public repository: \newline\href{https://github.com/nhurleywalker/GLEAM-X\_Periodic\_Transient}{https://github.com/nhurleywalker/GLEAM-X\_Periodic\_Transient}. \Fig~\ref{fig:LnuW} was generated using \newline\href{https://github.com/nhurleywalker/Transient\_Phase\_Space}{https://github.com/nhurleywalker/Transient\_Phase\_Space}. Further code can be supplied by the authors on reasonable request.
   \item[Correspondence] Correspondence and requests for materials should be addressed to N.H.-W. (nhw@icrar.org)

\end{addendum}

\section*{Extended Data}

% See http://s3-service-broker-live-19ea8b98-4d41-4cb4-be4c-d68f4963b7dd.s3.amazonaws.com/uploads/ckeditor/attachments/7823/3h_Extended_data.pdf for an explanation of this formatting

\renewcommand\thefigure{\arabic{figure}} 
\setcounter{figure}{0}
\renewcommand{\figurename}{Extended Data Figure}

\begin{figure}[H]
\includegraphics[width=0.5\textwidth]{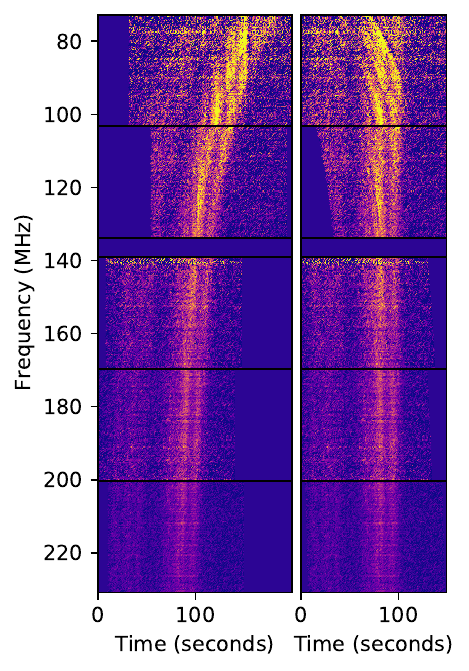}
\caption{Dynamic spectra of the observations used to measure the dispersion measure. From top to bottom, separated by horizontal black lines we show the observations 1205008192 (72--103\,MHz), 1205007112 (103--134\,MHz), 1205011432 (139--170\,MHz), 1205010352 (170--200\,MHz), and 1205009272 (200-231\,MHz). These observations were taken on 2018-03-13 between 20:12 and 21:24 (see \Fig~\ref{fig:light_curves}). The left panel shows the data aligned using a period of \srcPlong\,s, while the right panel shows the same, including a dispersion correction of \srcDM\,\DMunit. Strong ionospheric scintillation is visible in the 72--103\,MHz data, causing ripples in the brightness of the source over time.
\label{fig:dm}}
\end{figure}

\begin{figure}[H]
    \centering
    \includegraphics[width=\textwidth]{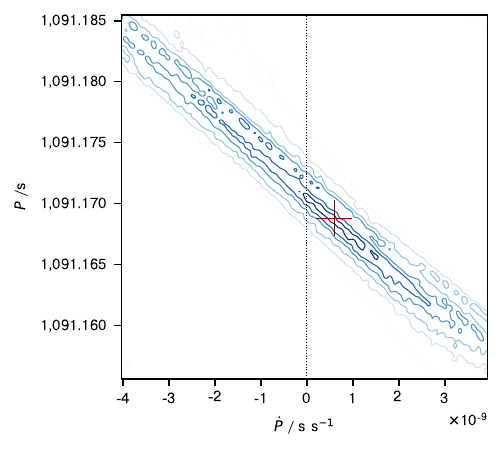}
    \caption{The explored search space in $P$ and $\dot{P}$ for the pulses recorded from \src{}. The contours show the peak flux density of the mean profile at 154\,MHz recovered at each combination of $P$ (vertical axis) and $\dot{P}$ (horizontal axis), in levels of 15, 14, 13, 12, and 11\,Jy. The best-fit values of $P=\srcPlong{}$\,s and $\dot{P}=\srcPdot{}$\,\Pdotunit{} are marked with a dark red ``$+$''.} 
    \label{fig:ppdot}
\end{figure}

\begin{figure}[H]
    \centering
    \includegraphics[width=\textwidth]{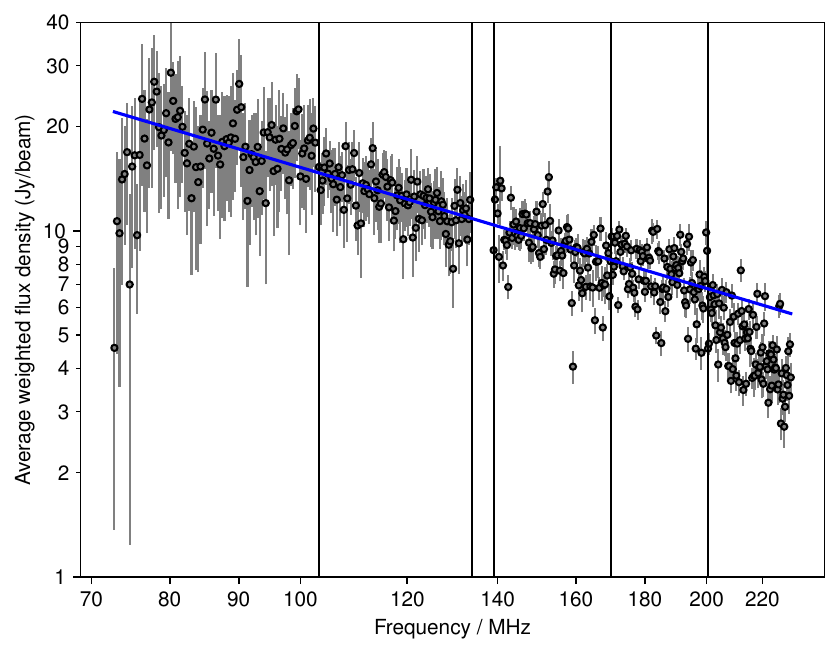}
    \caption{The flux density of \src{} as a function of frequency. This is derived from the same observations shown in \EFig~\ref{fig:dm}. Points are determined via an average of the source profile in each frequency bin, weighting by the signal-to-noise of the frequency-averaged profile. A power-law fit using the data spanning 95--195\,MHz is shown in blue, with $\alpha=\srcalpha\pm\srcalphaerr$.} 
    \label{fig:spectrum}
\end{figure}

\begin{figure}[H]
    \centering
    \includegraphics[width=\textwidth]{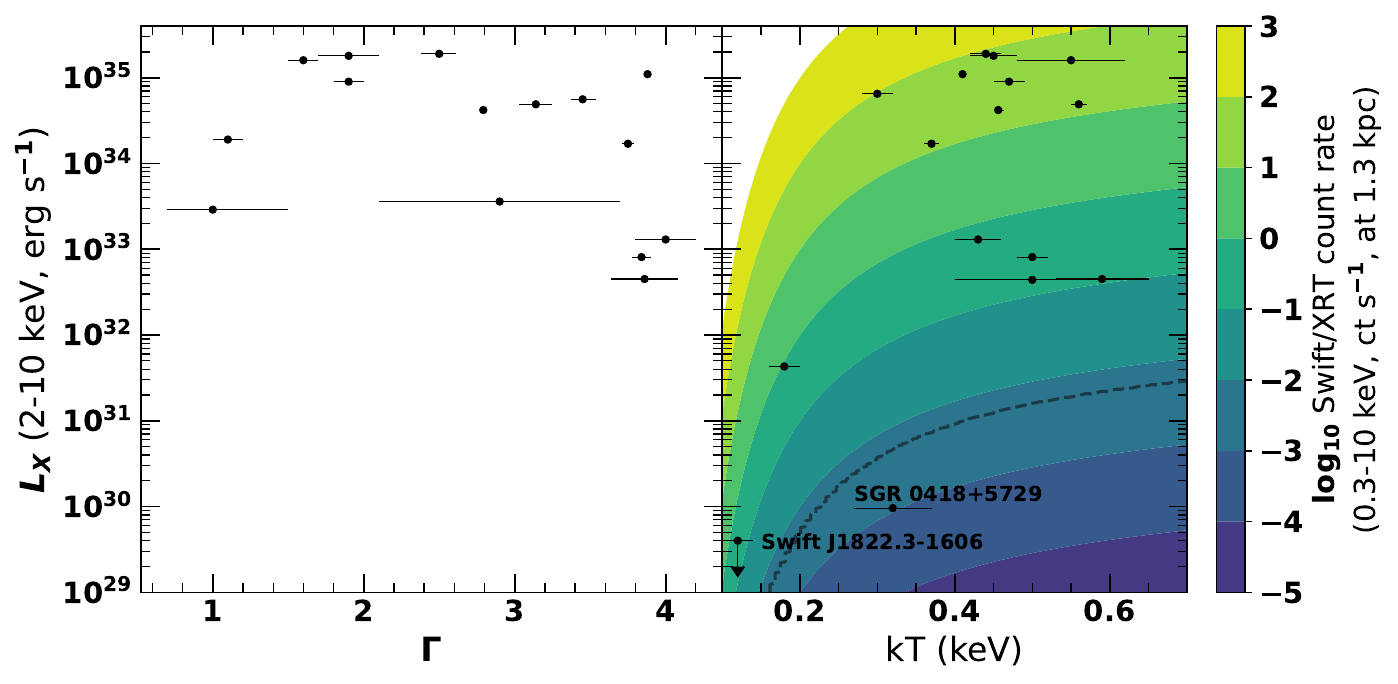}
    \caption{X-ray luminosity and spectral properties of magnetars\cite{2014ApJS..212....6O} compared to the X-ray luminosity limits of \src{}. The two faintest known magnetars are labelled. The colored contours represent expected {\it Swift}/XRT count rates for putative X-ray luminosities and blackbody temperatures for a source at 1.3\,kpc, assuming a hydrogen column density of N$_\text{H}\approx2\times10^{21}$ cm$^{-2}$. The gray dashed line represents the implied luminosity upper limit for \src{} based on the 3-$\sigma$ upper limit obtained from the {\it Swift}/XRT observation. From the magnetar fundamental plane, we predict \src{} to have a quiescent luminosity 4.5~orders of magnitude lower than our limit from {\it Swift}/XRT.} 
    \label{fig:xray_magnetars}
\end{figure}

\begin{figure}[H]
    \centering
    \includegraphics[width=\textwidth]{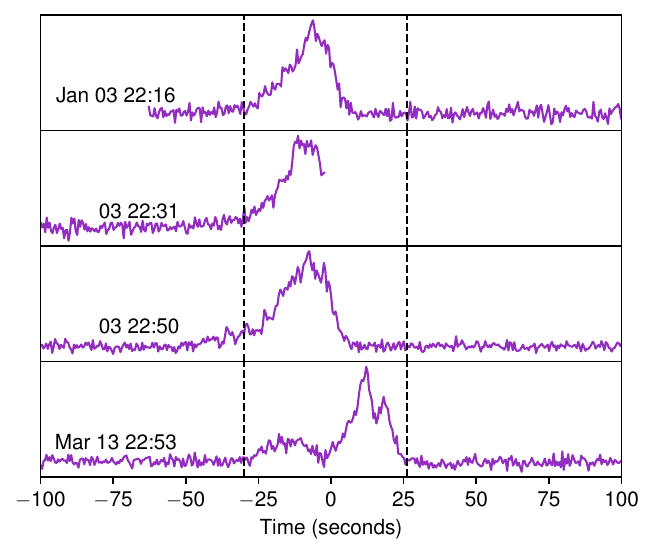}
    \caption{Pulse profiles for the three detections on Jan~03, compared with a wide pulse detected on Mar~14. Barycentric corrections and dedispersion have been applied. The data are all taken at the same frequency, 170--200\,MHz. Vertical dashed lines encapsulating the profile found on Mar~13 are overplotted to guide the eye.} 
    \label{fig:pulse_alignment}
\end{figure}

\begin{figure}[H]
    \centering
    \includegraphics[width=0.8\textwidth]{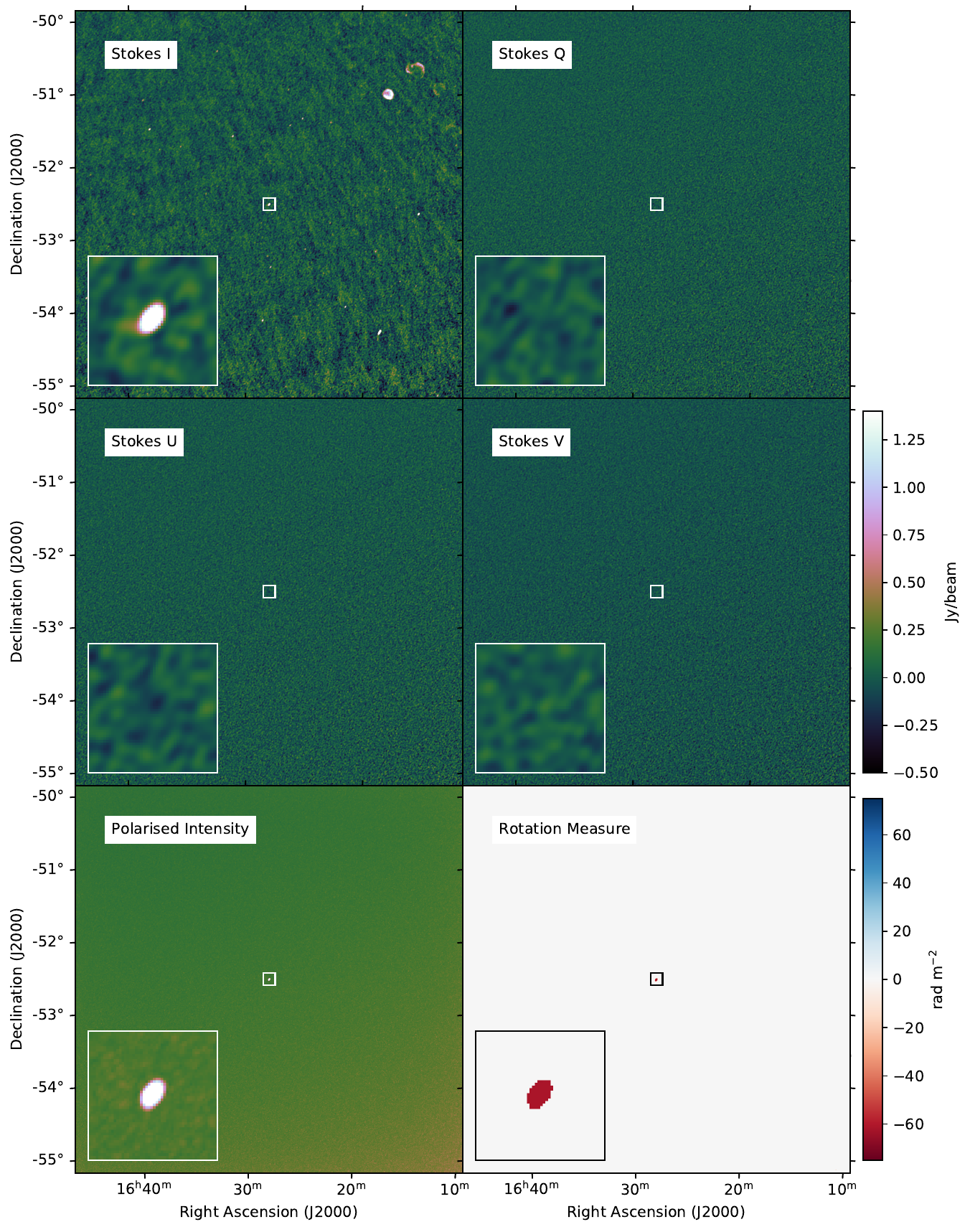}
    \caption{Images in full Stokes and polarised intensity (in Jy\,beam$^{-1}$), and rotation measure (in rad\,m$^{-2}$), of the region $5^\circ\times5^\circ$ around \src{}. The images were made using observation 1200354592 at 2018-01-18T23:49, using only the interval where the source was producing emission. Faraday rotation over the imaged bandwidth of 30\,MHz causes the Stokes Q, U, and V emission to average to zero. The polarised intensity shows the maximum value of the Rotation Measure spectrum. Where the polarised intensity is $<7\times$ the local noise, the corresponding RM value has been masked.} 
    \label{fig:iquvprm}
\end{figure}

\begin{figure}
\includegraphics[width=\textwidth]{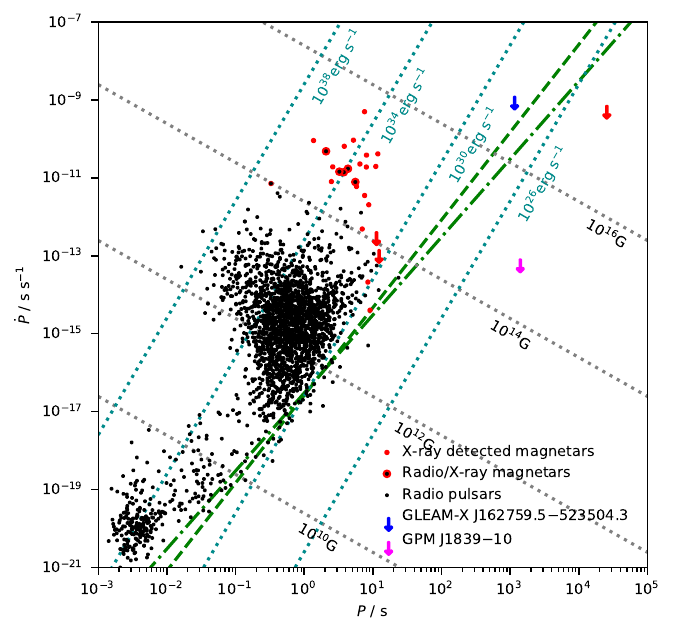}
\caption{A scatter plot of period derivative $\dot{P}$ against period $P$. \src{} is shown as a blue arrow with an upper limit on $\dot{P}$ (see Methods) in context with the known pulsars\cite{2005AJ....129.1993M} (black dots), X-ray-detected magnetars\cite{2014ApJS..212....6O} (red dots and arrows), and magnetars known to emit in both X-ray and radio (red circles around black dots).
The slowest (and radio-quiet) X-ray magnetar 1E\,161348$-5$055 is also shown with an upper limit on $\dot{P}$.
The green dashed and dot-dashed lines correspond to the theoretical ``death lines'' for pulsar radio emission for cases~I and III calculated by Zhang et al.\cite{2000ApJ...531L.135Z}.\label{fig:p_pdot}}
\end{figure}

%for arXiv

\begin{methods}

\subsection{Observations}
The Murchison Widefield Array is a low-frequency radio telescope operating in the Murchison region of Western Australia\cite{2013PASA...30....7T, 2018PASA...35...33W}. At the time of the observation in which \src{} was discovered, it was engaged in observing the GaLactic and Extragalactic All-sky MWA -- eXtended (GLEAM-X) survey, a follow-up to the GLEAM survey\cite{2015PASA...32...25W}. The data were taken in a drift scan mode, iterating through a 72--231\,MHz bandwidth by dwelling for 2\,minutes in each of five 30.72-MHz bands, yielding a resolution of $5'$--$45''$ and snapshot noise levels of $150$--$25$\,mJy\,beam$^{-1}$. Over this bandwidth the field-of-view of the instrument is $60^\circ$--$25^\circ$ across, yielding multiple measurements of sources drifting through the primary beam. The data are sampled at 0.5-s time resolution and 10-kHz frequency resolution. We downloaded measurement sets from the MWA All-Sky Virtual Observatory (https://asvo.mwatelescope.org/) and at this stage averaged the data to 40\,kHz to reduce data size and decrease processing times. 

The initial detection of the transient was made by performing a differencing of the visibilities between observations taken at an identical Local Sidereal Time several months apart (Hancock et al. in prep). Two detections were made, and based on the non-detections in adjacent observations, an initial period was determined. Including a barycentric correction, and searching the archive thoroughly between July~2017 and July~2018, the interval of activity was established to be January~2018 to March~2018, with a total of \ndetections{}~detections (see Supplementary Methods for further details on searching the MWA archive).

\subsection{Calibration and imaging}
For all data, calibration was performed using the \textsc{mitchcal} algorithm\cite{2016MNRAS.458.1057O} to derive antenna gains by comparing the raw data to model visibilities formed from a sky model based on catalogues derived from GLEAM\cite{2017MNRAS.464.1146H, 2019PASA...36...47H}. These were applied to the observation, and deep imaging was subsequently performed using \textsc{WSClean}\cite{2014MNRAS.444..606O} version~2.9.0, creating a visibility model that includes the spectrally variant primary beam. A $10'\times10'$ region around \src{} was masked during this process, so that the source is not included in the model. 

After the deep model was created and subtracted from the visibilities, the data were re-imaged such that the transient was the only source in the field. The data were imaged at 320\,kHz frequency resolution and 0.5\,s time resolution in full Stokes (I,Q,U,V), which were formed from the instrumental data by application of the MWA beam model\cite{2015RaSc...50...52S}. We also used observations of a polarisation calibrator (PKS~J0636$-2036$) to correct radio emission leaking from Stokes~U into Stokes~V (X-Y phase calibration\cite{2017PASA...34...40L}).

%Since strong pulse profile evolution is observed, to determine the dispersion measure, we used the five observations which spanned 72 -- 231\,MHz in the shortest interval of time, 90~minutes (\Fig~\ref{fig:dm}). 
%Recommended rephrasing
\subsection{Dispersion Measure}
Since strong pulse profile evolution is observed, we used the five observations that spanned 72--231 MHz in the shortest interval of time (74~minutes; \EFig~1) to calculate the dispersion measure.
We aligned these observations using a period of \srcPlong\,s, then ran dedispersion trials in steps of 0.5 over range 50 to 60\,\DMunit, finding that $\srcDM\pm\DMerror{}$\,\DMunit{} produced the best fit. Light curves (\Fig~2)
%\ref{fig:light_curves})
were then produced for each observation by applying dedispersion and averaging the dynamic spectra along the frequency axis.

\subsection{Polarisation}
Polarisation analysis was performed following the methods of the POlarised GLEAM Survey (POGS\cite{2018PASA...35...43R,2020PASA...37...29R}). From all observations with detections, only observations within 170--230 MHz were selected for the polarisation analysis to avoid depolarisation caused by the (40\,kHz) channel width. We performed RM synthesis\cite{brentjens2005faraday} on the time-averaged Q/U spectrum of each observation with the RM Tools software (\href{https://github.com/CIRADA-Tools/RM-Tools}{https://github.com/CIRADA-Tools/RM-Tools} to obtain the corresponding RM, fractional polarisation, and polarisation angle. To investigate any variation of the polarisation angle within the pulse phase, we also performed RM synthesis on the Q/U spectrum of each timestep for two high S/N observations. We found that the polarisation angle was constant with respect to time within and between observations (e.g. the middle two panels of \Fig~3).
%\ref{fig:dynspec}).

\subsection{Period and Period Derivative}

We used phase dispersion minimization based on Lafler-Kinman statistic\cite{1965ApJS...11..216L,1978ApJ...224..953S,1996ApJ...460L.107S,2002A&A...386..763C} to quantify the periodicity of the pulses, using the package \textsc{p4j}\footnote{\url{https://github.com/phuijse/P4J}}. This choice is motivated by the variable shape and amplitude of the pulses and presence of large irregular gaps in data (which occasionally truncate part of an observed pulse). Using this method we found a clear peak in the periodogram with a period of $1091.170$\,s.

The period obtained from the periodogram analysis above assumes that the period is constant (i.e. $\dot{P} = 0$).
In order to place constraints on $\dot{P}$, we performed a grid search of $P$ and $\dot{P}$ values, centred on $1091.170$\,s and $0$\,\Pdotunit{} respectively, to find which pairs of values are consistent with the observed arrival times of the pulses.
We searched over the ranges $1091.150$\,s$\le P \le 1091.185$\,s and $-4\times10^{-9} \le \dot{P} \le 4\times10^{-9}$, aligning the pulses in pulse phase for each value pair, and taking the peak flux density of the averaged pulses (the ``mean profile'' in pulsar parlance) as our metric for goodness of alignment.
Since different pulses were observed in different frequency bands, each brought to a common frequency of 154\,MHz using $\alpha=\srcalpha{}$ (see below), before averaging.

As shown in \EFig~\ref{fig:ppdot}, the resulting $P$ and $\dot{P}$ were somewhat degenerate, as expected due to the three-month time window of measurements and the evolving pulse profile. The maximum peak flux density 15.5\,Jy corresponds to the values $P = \srcPlong{}\pm\srcPerr{}$\,s and $\dot{P} = \srcPdot{}$\Pdotunit{}. The most significant contour in \EFig~\ref{fig:ppdot} is shown at 15\,Jy to show that there are multiple values of $P$ and $\dot{P}$ which produce a similar flux density. Therefore, this analysis favours the ranges $0 < \dot{P} < 1.2\times10^{-9}$, i.e. spin-down is more likely than spin-up. Note that the error on the source period is representative for any choice of $\dot{P}$ within this range, reflecting the width of the most significant contour in the period dimension.

\subsection{Spectral index}
\label{sec:spectral_index}

Individual 30.72-MHz observations of \src{} lack the frequency coverage to determine an accurate spectral index, but since the flux density clearly varies with time, averaging over many observations is unlikely to yield a usable result. To obtain a good estimate, we used the same observations with which we measured the dispersion measure: 1205008192 (72--103\,MHz), 1205007112 (103--134\,MHz), 1205011432 (139--170\,MHz), 1205010352 (170--200\,MHz), and 1205009272 (200-231\,MHz); these are the observations taken on 2018-03-13 between 20:12 and 21:24 (\Fig~
%\ref{fig:light_curves}
2), where the pulse profile is reasonably consistent with time. After dedispersion and alignment on the source period, we obtained an average profile by averaging the data along the combined frequency axis (vertical axis in the right-hand panel of \EFig~\ref{fig:dm}). From this we selected the timesteps where the source was clearly ``on'' (timesteps 0--100 in the right-hand panel of \EFig~\ref{fig:dm}), and for each frequency, determined the weighted average flux density, using the average profile as a weighting function.

Errors were calculated as the root-mean-squared noise of each slice for unflagged timesteps during which the source was ``off'', divided by the square root of the number of samples in the average flux density measurement, added in quadrature with a 5\,\% flux density calibration error. \EFig~\ref{fig:spectrum} shows the weighted averages and their errors plotted as a function of frequency. The drop-off at low frequencies is likely due to averaging over the strong ionospheric scintillation (time-dependent striping visible in the top of \EFig~\ref{fig:dm}), while the steepening at the higher frequencies may be intrinsic, but the signal-to-noise is low so it is difficult to be sure. Therefore, we used the measurements taken between 95 and 195\,MHz and the \texttt{scipy} implementation of a the Levenberg-Marquardt least-squares fit to determine a radio spectral index of $\alpha=\srcalpha\pm\srcalphaerr$, where the flux density $S_\nu \propto \nu^\alpha$. The reduced-$\chi^2$ of the fit is 1.86.

\subsection{Error on the distance estimate}

Models of the Galactic electron density can be used to convert the dispersion measure (DM) into a distance. YMW~2016, the most recent model\cite{2017ApJ...835...29Y}, derives a relative distance error $\frac{D_m - D_i}{D_i}$ using 189~pulsars for which distances are known: $D_m$ is the model distance based on the observed DM and $D_i$ is the independently determined distance (e.g. by parallax). For their sample, the root-mean-square of the relative distance error is 0.398. From version 1.65 of the Australia Telescope National Facility Pulsar Catalogue\cite{2005AJ....129.1993M}, we extracted data for the five pulsars with independently-measured distances within $10^\circ$ of \src{}. For these pulsars, we found that the root-mean-square of the relative distance error is 0.393. Thus we conclude that there is a $\sim40$\,\% error on the estimated distance of 1.3\,kpc.

\subsection{Position measurement}
At radio frequencies, the apparent source position measured by an interferometer may be shifted by an angular offset $\Delta \theta$ by the ionosphere, proportional to the transverse gradient $\nabla_{\perp}$ of the total electron content (TEC) toward the source and the square of the wavelength at which the observations were taken.
% Removed \cite{TMS} because we have already hit the reference limit
To determine an accurate position for \src{}, we used observation 1205009272, taken at the highest frequency band, 200 -- 231\,MHz, on a quiet night during which ionospheric distortions were minimised, and imaging just the 30-s subset of the observation during a high S/N pulse. We used the software \textsc{fits\_warp}\cite{2018A&C....25...94H} to calculate the local position shift based on the apparent shifts of nearby compact bright calibrator sources. The error was calculated as the mean residual offset of these nearby sources after the shift had been modelled and removed, i.e. $2''$. The derived position is R.A. $16^\mathrm{h}27^\mathrm{m}59.5^\mathrm{s}$, Dec. $-52^\circ35'04.3''$.

%\begin{equation}
%\Delta \theta = - \frac{1}{8\pi^2}\frac{e^2}{\eta_0 m_e } %\frac{1}{\nu^2} \nabla_\perp \mathrm{TEC}
%\label{eq:d_theta}
%\end{equation}

\subsection{\textit{Swift}/XRT Observations}
We observed \src{} with the Neil Gehrels {\it Swift} observatory\cite{2004ApJ...611.1005G} for 2\,ks with the X-ray telescope (XRT)\cite{2005SSRv..120..165B} in photon-counting mode.
%and the UV/optical telescope (UVOT)\cite{2005SSRv..120...95R} with uniform weighting in all 6 filters.
The observations were taken from 2021-02-25 16:34:00 to 2021-02-25 21:27:00 (UTC) and the observation id was 00014085001.
We reduced and analyzed the data using HEASOFT~v6.28\cite{2014ascl.soft08004N}, and the \textsc{xrtpipeline}~v0.13.5. 
There is no detection of a point source in any bands, nor is there any indication of extended structure.
We estimate a 3$-\sigma$ upper limit of 5.6$\times10^{-3}$\,ct\,s$^{-1}$ for the count rate in the 0.3--10\,keV band for \src{}. Assuming a thermal spectrum (blackbody, $kT=0.1$\,keV), the upper limit in count rate corresponds to an absorbed flux of 1.5$\times10^{-13}$ \ergpers\,cm$^{-2}$ in the 0.3--10\,keV band.
Assuming a non-thermal spectrum ($E^{-\alpha}$; $\alpha=2$, typical for magnetars\cite{2014ApJS..212....6O}), the upper limit in count rate corresponds to an absorbed flux of 1.9$\times10^{-13}$\,\ergpers\,cm$^{-2}$ in the 0.3--10\,keV band. \EFig~\ref{fig:xray_magnetars} shows the corresponding luminosity limit calculated as a function of $kT$.

% Left out since Arash doesn't have time to check
%\subsection{Optical analysis}
%The Dark Energy Survey (DES) (TODO Arash add ref) covers the region of \src{} in $grziY$ filters (\Fig~\ref{fig:optical}). We inspected the two sources within our positional error ellipse (A and B) to determine whether they could be an optical counterpart to the observed radio emission. We also examined the more distant sources C, D, and E to check our measurements. The unlabelled source is too confused for accurate photometry. Examining the spectral energy distributions of the five sources, we find that they are consistent with stellar continua. The DES data were observed in 2016, 2017, and 2019; no astrophysical variability was found for any of these sources. We also used the Gaia atlas to determine distances to these sources (TODO ARASH REF), and find that they are $>3$\,kpc away, inconsistent with the distance measured for \src{}. Proper motions of all of these sources as measured by Gaia are consistent with normal stars.

\subsection{Radio luminosity calculation}

Working under the assumption that \src{} is a pulsar or magnetar in order to determine its energetics, we can transform the observed flux density into a radio luminosity.
To precisely determine the radio luminosity of a rotating magnetic neutron star its geometry with respect to the observer needs to be known; this is often derived by examining the change in pulse phase with respect to time.
In the case of \src{}, the pulse phase is flat, similar to the case of the radio magnetar Swift~J\,$1818.0-1607$\cite{2020MNRAS.498.6044C}. In this case, we cannot derive the geometry of the emission cone
%\cite{2021MNRAS.502.1549T}
and instead interpret the flat phase as our line-of-sight just grazing the edge of the emission cone (i.e. the impact angle between the line-of-sight and the magnetic axis is similar to the emission cone opening angle). This is qualitatively consistent with the pulse duty cycle of 30--60\,s of activity in \srcP{}\,s, $\approx3$--6\,\%. For pulsars with this duty cycle, for typical opening angles of $6^\circ$, the radio luminosity at 1.4\,GHz is\cite{2012hpa..book.....L} $7.4\times10^{30}\frac{D}{\mathrm{kpc}}^2\frac{S_\mathrm{1.4GHz}}{\mathrm{Jy}}$\,\ergpers. \src{} produces pulses of peak flux densities of up to $S_\mathrm{154MHz}=45$\,Jy. Scaling this to 1.4\,GHz by $\alpha=\srcalpha{}$ we would expect $S_\mathrm{1.4GHz}=3.5$\,Jy, and therefore $L_\mathrm{1.4GHz}=4\times10^{31}$\ergpers.

\subsection{Long-term duty cycle}
In 8~years of operation, the MWA has accumulated $\sim160$\,hours of observing time that have pointing directions within 15$^\circ$ of \src{}, and might in principle be sensitive to it. However, since the data span many different projects, array configurations, frequencies, and observing modes, and processing data takes about $100\times$ longer than observing, searching this data thoroughly is a daunting task. Additionally, only when the data are taken in a contiguous $\sim$20-minute block may emission from this source be ruled out. We first examined the data before the 2018-Jan -- Mar activity window using the period of the source, and found no emission from \src{} in 2017-Dec or 2018-Apr. We searched five blocks of 20-minute contiguous observations in 2017-May, 2017-Oct, 2017-Nov, 2018-Apr, 2018-May, finding no detections. Before 2017-May and after 2018-Jun, the telescope was reconfigured into compact ``Hex'' mode, a redundant-baseline configuration with very poor imaging quality, yielding poor constraints from data taken in that mode. We searched for suitable observations in the archive as early as possible (2014-Mar and 2014-Jun), and also took new observations (2021-Feb), finding no detections. There remains in the archive $\sim15$\,hours of data in suitable contiguous 20-minute blocks that may be searched in a future paper.

At the time of writing, the MWA has been in operation for 8~years, and this source was only found to be active for 2~months in that time, yielding an estimate of the duty cycle of $\frac{60\,\text{days active}}{3000\,\text{days searched}} = 2$\,\%. Alternatively, we could assume that the source was active any time that we have not thoroughly searched the data, and with the co-operation of a conspiratorial Universe, the duty cycle can trend toward 100\,\%. Or, this might be the only time in the (unknown) lifetime of the source that it has produced emission, resulting in a duty cycle trending toward 0\,\%. It is also worth noting that the source may not be entirely inactive during our non-detections, but its pulsations may be below our detection threshold; follow-up with a more sensitive radio telescope would be illuminating. For this work, we use the estimate of a $2$\,\% long-term duty cycle for pulses above the detection threshold of the MWA.

\subsection{Pulse profile evolution and alignment}

At first examination, the three pulses recorded on Jan~03 appear to be misaligned with the other pulses. However, the pulses recorded on Mar~13 show an early ``secondary'' pulse preceeding the main pulse by $\sim15$\,s. Aligned by our determined $P$ and $\dot{P}$, the pulses from Jan~03 fit within the wide pulse profiles found in Mar~13 (\EFig~\ref{fig:pulse_alignment}). We therefore suggest that there is no misalignment, and instead the apparent effect is caused by pulse profile evolution.

\end{methods}

\end{document}

%\newpage
%\section{Supplementary Methods}

%\section{Supplementary Equation(s)}

%\section{Supplementary Notes}
%(including notes clarifying statistical analyses, acknowledgements, grant or other numbers)}

\section{Software}

 We acknowledge the work and support of the developers of the following following python packages: Astropy\cite{TheAstropyCollaboration2013}, Numpy\cite{vaderwalt_numpy_2011}, and Scipy\cite{Jones_scipy_2001}.
 We also made extensive use of the visualisation and analysis package DS9 (\href{http://ds9.si.edu/site/Home.html}{http://ds9.si.edu/site/Home.html}).
 %and Topcat \cite{Taylor_topcat_2005}. 
 This work was compiled in the very useful free online \LaTeX{} editor Overleaf.

\bibliography{refs.bib}